\documentclass{iaus}
\usepackage{graphicx}

\title[JD 11.~~Chemical abundances of Extragalactic PNe] 
{Extragalactic Planetary Nebulae: \\tracers of the chemical evolution of nearby galaxies}
 \author[L. Magrini, L. Stanghellini, D. R. Gon\c calves]   
{Laura Magrini$^1$,
Letizia Stanghellini$^2$,
Denise R. Gon\c calves$^{3,4}$
}

\affiliation{$^1$INAF-Osservatorio Astrofisico di Arcetri, Largo E. Fermi 5 \\
I-50125, Firenze, Italy \\ email: {\tt laura@arcetri.astro.it}\\[\affilskip]
$^2$National Optical Astronomy Observatories, Tucson, AZ 85719\\ email: {\tt lstanghellini@noao.edu}\\[\affilskip]
$^3$UFRJ - Observat\'orio do Valongo, La. Pedro Antonio 43, 20080-090 Rio de Janeiro, Brazil\\[\affilskip]
$^4$Department of Physics and Astronomy, University College London, Gower Street, WC1E 6BT  London, UK\\ email: {\tt denise@astro.ufrj.br}
}

\pubyear{2012}
\volume{283}  
\pagerange{1--8}
 \jname{Planetary Nebulae: An Eye to the Future}
\editors{A. Manchado, L. Stanghellini \&  D. Sch\"onberner, eds.}
\begin{document}

\maketitle

\begin{abstract}
The study of the chemical composition of Planetary Nebulae in external galaxies is of paramount
importance  for the fields of stellar evolution and chemical enrichment history of galaxies.
In recent years a number of spectroscopic studies with 6-8m-class telescopes have been devoted
to this subject improving our knowledge of, among other, the time-evolution of the radial metallicity
gradient in disk galaxies, the chemical evolution of dwarf galaxies, and stellar evolution at
low metallicity.
\end{abstract}

\firstsection 
\section{Introduction}

The determination of the chemical composition of Planetary Nebulae (PNe) in external galaxies is extremely
important both from the point of view of the stellar evolution, allowing us to study the stellar
nucleosynthesis in different conditions, and from that  of
the formation and evolution of galaxies, giving us the unique possibility to determine directly
the chemical composition of the interstellar medium (ISM) in different epochs that the present one, such as, for example, at the time of the
formation of the PN progenitors.

The two points of view described above are strictly related. In fact,  to understand  the chemical evolution of the
host galaxy, we need to constrain the ISM chemical history, which in turn can be done effectively by knowing the elemental abundances of evolutionary-invariant
elements. It is thus necessary to understand where and in which conditions (e.g., mass range,
metallicity) elements such as oxygen or neon could be dredged-up to the stellar surface and consequently
modify the composition of the original interstellar cloud, where the progenitor star did form.

In this review we present recent results obtained with large-aperture, multi-object  spectroscopy of extragalactic PNe.
section 1 analyzes the PN and HII region abundances, and their radial metallicity distribution, in the spiral galaxies M33 and M81. The two populations
allow inspection of 
time-evolution of the radial metallicity gradient. 
In Section 2 we present the properties of PNe in dwarf galaxies, and their impact on chemical evolutionary models. 

\section{The time-evolution of the radial metallicity gradient}

One of the most discussed questions about the radial metallicity gradient in
disk galaxies is its evolution with time, which to date cannot be constrained by the existing arrays of chemical evolutionary models.
Classical chemical evolutionary models predict different temporal behaviors depending
on assumptions over the  gas inflow and outflow rates, the star and cloud
formation efficiencies, and other set-up physical conditions. Models might predict a steepening
of metallicity gradients with time (e.g., Chiappini et al. \cite{chiappini97}, \cite{chiappini01}), or, conversely,  a flattening 
of such gradients (e.g., Moll{\'a} et al. \cite{molla97}, \cite{molla05}; Magrini et al. \cite{magrini07b}).
The main differences between the two groups of models are the efficiency of the enrichment processes in the inner and outer regions of the disk,
and the nature of the material (primordial or pre-enriched) falling from the halo onto the disk.
In all the above-cited models, however, the dynamical effects, and the possible migration of stellar populations, have been neglected, thus they might oversimplify
the actual problem under exam (see for example  L{\'e}pine et al (\cite{lepine11}) for a description of the effect
of spiral arms on gradients).

To better understand galaxy evolution we need observational constraints for the 
theoretical scenarios.
The comparison between chemical abundances of PNe and  HII regions seems to be a very promising way to
set strong constraints to the models, since the elemental abundances of both populations can be derived using the same set of observations, 
the similar data reduction and
analysis techniques, and identical abundance determination methods, yet they define different evolutionary times of the galaxy. In this way it is possible to avoid most of the biases due to the 
stellar vs. nebular analyses, in particular
allowing the direct comparison of the same element, oxygen, at two epochs, and thus avoiding the dangerous  comparison between $\alpha$- and
iron-peak-elements, whose ratio depends on  the star formation history of the galaxy.

\subsection{The PN populations in M33 and M81}

We have studied the PN and HII region populations in two  nearby spiral galaxies, M33 and M81.
M33 (also called NGC~598) belongs to the Local Group, at the distance of 850~kpc (Galleti et al. \cite{galleti04}, Sarajedini et al. \cite{sara06}).
The M33 metallicity gradient derived from $\alpha$-element abundances has been
the focus of several studies, involving both PNe and HII regions (e.g., Crockett et
al.~\cite{crockett06}, Magrini et al. \cite{magrini04}, \cite{magrini07a}, \cite{magrini09}, \cite{magrini10}, Rosolowsky \& Simon \cite{rs08}, Rubin et al. \cite{rubin08}, Bresolin \cite{bresolin11}).
On the other hand, M81 (also NGC~3031) is the largest member of the nearest interacting group of
galaxies at a distance of 3.60~Mpc (Freedman et al. \cite{freedman01}).
Until a recent spectroscopic study (Stanghellini et al. \cite{st10}), the $\alpha$-element abundances of PNe and HII regions in M81
have remained elusive, with the exception of 17 H II regions studied by
Garnett \& Shields (\cite{gs87}),
The advantage of studying M33 and M81 instead of the Galaxy as
a playground for gradients is that extragalactic PNe have
well determined galactocentric distances (with relative errors within
5$\%$), compared to the large indetermination of Galactic PN distances
(Stanghellini et al.~\cite{stanghellini08}).

\begin{figure}
\includegraphics[width=12cm]{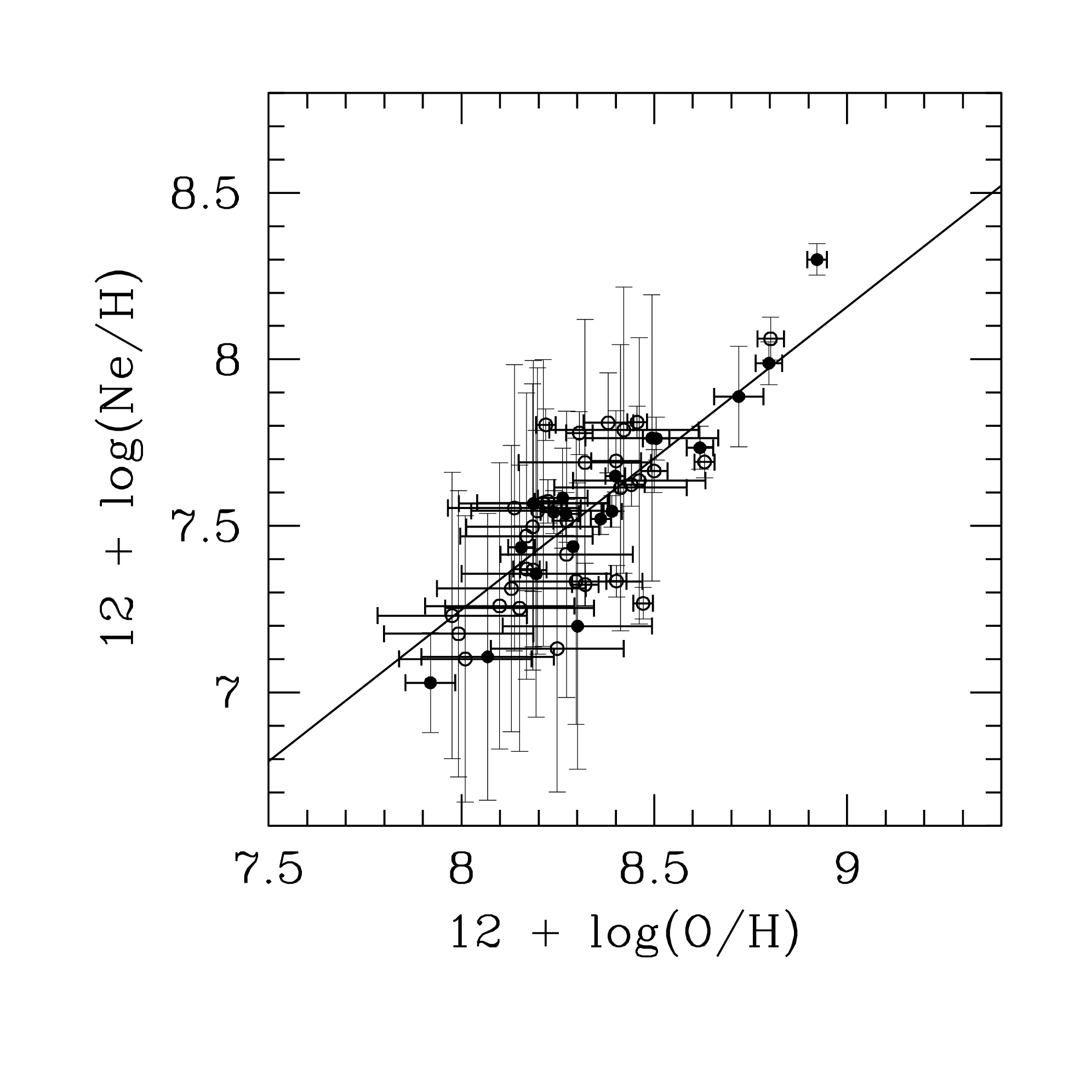}
\caption{The relationship between oxygen and neon abundances in M33. Type I PNe and non-type I PNe
are plotted as filled circles and empty circles respectively, according to the definition of
Dopita
\& Meatheringham (\cite{dopita91}).
The continuous line is the weighted least square fit to the complete sample of PNe.}
\label{Fig_oxyne}
\end{figure}

The first step of the gradient analysis is to confirm which elements can be used to trace the past ISM composition.
The best candidates are oxygen and neon, and we should find out whether
these elements could be modified through 
low- and intermediate-mass  stars (LIMS) evolution.
A possible way to verify this is to study the relationship between neon
and oxygen abundances.  These elements derive both from primary
nucleosynthesis, mostly in stars with M$>$10 M$_{\odot}$.  If the O/H
and Ne/H abundances are
really independent of the evolution of the PN progenitors through the
AGB phase, we should observe a tight correlation between their abundances.
In Figure~\ref{Fig_oxyne} we plot O/H against Ne/H for the 55 M33 PNe where both
abundances were available.  The
slope of the correlation is close to unity, 0.90$\pm$0.11, pointing at a locked variation of
these two elements, and confirming  that there is no
evidence of enhancement of oxygen and neon in PNe at M33 metallicities.
It is possible that nucleosynthesis could affect the alpha-element abundances at much lower metallicities. But this effect would be null at M81
metallicities, higher than those of M33, thus for both galaxies analyzed here we could safely assume O vs. Ne evolutionary lockstep (Stanghellini et al.
2010). 

The second step is to identify the PNe with the least massive progenitors, thus the oldest ones, such as for example the non-Type I PNe. In this way one
could detect the signature of galactic chemical evolution in the comparison of old PN progenitors with the young stellar population traced by the HII regions.
If we examine the locus of M33 and M81 disk PNe on the He/H-N/O plane we can single out the Type I PNe, and then select the remaining PNe as the oldest
population in either galaxy. It is worth recalling that the Type I PNe are metallicity dependent (the amount of
nitrogen produced by hot bottom burning is dependent on
the amount of carbon, and the oxygen abundance
depends on the metallicity of the galaxy, see Magrini et al. \cite{magrini09} and Stanghellini et al. \cite{st10} for details).
No  Type I PN has been detected in M81. Also, interestingly, no M81 PN shows the HeII 4686 emission line.
On the other hand, about 20\% of the observed PNe in M33 are of Type I, and we detected HeII 4686 in many PNe.
We conclude that the populations of the brightest PNe in galaxies strongly depend upon galactic metallicity. 

In fact, at high metallicities, such as in M81, the brightest PNe are not those with the hottest Central Stars (CSs).
The post-AGB shells with very hot CSs are enshrouded in dust at early phases of their evolution, thus still thick to the H$\alpha$ radiation.
The thinning time depends on metallicity, being shorter at low metallicity, and too long at M81 metallicity to allow us to observe PNe
with massive progenitors among the brightest ones. For this reason in  M81 we detect only PNe with low mass progenitors.

\subsection{The radial metallicity gradients}
\begin{figure}
\includegraphics[width=10cm,angle=-90]{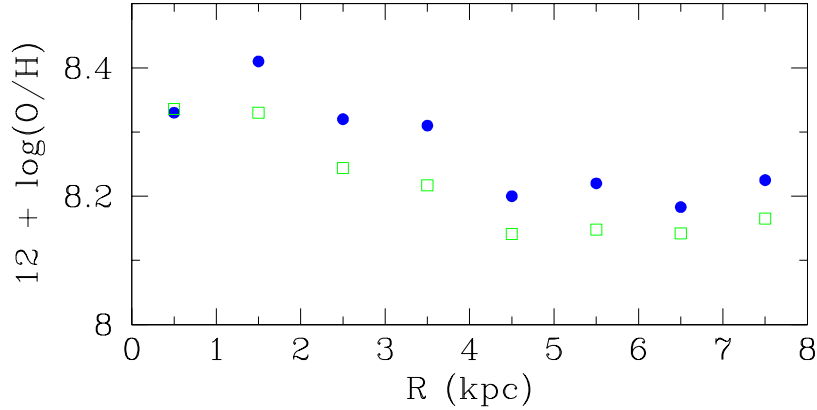}
\caption{Oxygen radial gradient of M33: HII regions averaged on bin of 1 kpc are shown with filled circles (blue), non-Type I PNe with empty squares (green).}
\label{fig_m33_grad}
\end{figure}

Once defined the sample of old PNe, we compared their radial gradient with that of HII regions.
In M33 we found for non-Type PNe (72 objects)
\begin{equation}
12+\log(O/H)=(8.41\pm0.06)-(0.030\pm0.013)\times {\rm R}_{GC}
\end{equation}
where R$_{GC}$ is the de-projected galactocentric distance.
 For HII regions, including the literature data described in Magrini et al. (\cite{magrini10}) and the new
 data of Bresolin (\cite{bresolin11}) for the HII regions located in the central part of M33,
 \begin{equation}
 12+\log(O/H)=(8.50\pm0.02)-(0.045\pm0.006)\times {\rm R}_{GC}.
\end{equation}
By taking into account the new data for  HII regions by Bresolin \cite{bresolin11},
there is a slight trend to a metallicity gradient steepening with time.
In addition we found that the global metal content has evolved $\sim$0.1 dex from the epoch of the formation of the PN progenitors to the present-time.
The result is more appreciable if we compare 12+$\log$(O/H) averaged on bins of 1 kpc as shown in Figure~2.

In M81 we can play with less objects, finding for PNe (19 objects)
\begin{equation}
12+\log(O/H)=(8.72\pm0.18)-(0.055\pm0.02)\times {\rm R}_{GC}
\end{equation}
and for HII regions (31 objects, from Garnett \& Shields \cite{gs87} and from Stanghellini et al. \cite{st10})
\begin{equation}
12+\log(O/H)=(9.37\pm0.24)-(0.093\pm0.02)\times {\rm R}_{GC}.
\end{equation}
The tendency seems again to favour a steepening with time, but the sample of HII regions for which a direct measurement
of the electron temperature is available is very limited both in number and galactocentric distance. The result for M81 need a further investigation
extending the analysis
to the very central regions and the extreme periphery of the galaxy.

It is very interesting to compare ther above findings with one of the most recent chemical evolutionary models (Rahimi et al.  \cite{rahimi11}).
These Authors derived the evolution of the radial abundance gradients of the disk stars of a
galaxy simulated with a three dimensional, fully cosmological chemical and dynamical
galaxy evolution code.
The comparison of their gradient for ÒyoungÓ disk stars. i.e. formed during the final 2 Gyr of the galaxy life-time,
with ÒintermediateÓ disk stars with ages  2~Gyr$<$Age$<$6~Gyr is very similar to the observational findings described above
for HII regions and non-Type I PNe: a translation toward higher metallicity, indication a global enrichment, and a
slight tendency to have gradient for the youngest population stepper than that of the oldest one.

\section{The chemical evolution of dwarf galaxies}

PNe belonging to dwarf galaxies are extremely interesting both because they allow us to investigate
stellar evolution at low metallicity,  and they give information
about chemical evolution of dwarf galaxies.
During the past five years the knowledge about PNe in dwarf galaxies has improved, and now
the physical and chemical parameters of PNe in many nearby dwarf galaxies have been derived. A recent reference list includes
the dwarf Irregular (dIrrs) galaxies Sextans A and B (Magrini et al. \cite{magrini05}, Kniazev et al. \cite{kniazev05}), IC~10
(Kniazev et al. \cite{kniazev08}, Magrini \& Gon\c calves \cite{mg09}, Yin et al. \cite{yin10}), and NGC3109 and NGC6822
(Pena et al. \cite{pena07}, Hern{\' a}ndez-Mart'nez et al. \cite{hm09}); and the dwarf spheroidal companions of M31, NGC 185, NGC205, M32, and NGC147
(Gon\c calves et al. \cite{goncalves07}, Richer \& McCall \cite{rm08}, Gon\c calves et al. \cite{goncalves11}) and of our Galaxy, Fornax and Sagittarius
(Kniazev et al. \cite{kniazev07}, Kniazev et al. \cite{kniazev08}).
In the following we describe some of the the most relevant results among the above papers.
\subsection{Third dredge-up and positive oxygen yields at low metallicity}

\begin{figure}
\includegraphics[width=12cm,angle=0]{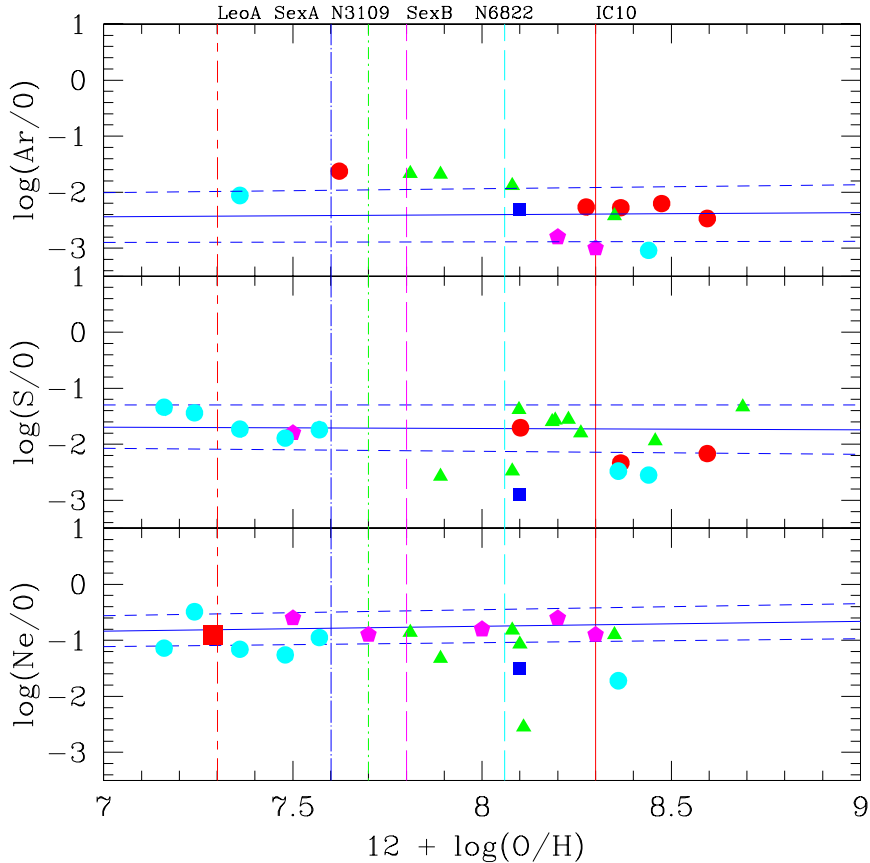}
\caption{ Ne/O, S/O and Ar/O vs. O/H.
Different symbols are for PNe in the different galaxies; large filled circles: IC10; filled triangles: NGC3109; large filled squares: Leo A; small filled
squares: Sextans A; pentagons: Sextans B ; small filled circles: NGC~6822. The nearly-horizontal lines show the abundance
line-ratios given by the a sampleof  HII regions in blue compact galaxies, as compiled by Izotov et al. (\cite{izotov06}).
The vertical lines show the average 12+log(O/H) locus of the samples of HII regions in each of the dIrr galaxies of this plot: lines
are labelled with the corresponding galaxy identification at the top of the plot.}
\label{fig_TDU}
\end{figure}

One important aspect of stellar evolution is whether or not the efficiency of
the third dredge-up (TDU, i.e., the process which brings the elements characterising the PN composition
to the stellar surface during the AGB phase) depends on metallicity.  Some stellar evolutionary models
claim that such a dependence is strong, and that when the metallicity is very low, even the elements previously not known to be dredged-up, such as neon
and oxygen, are indeed carried to the stellar surface (e.g. Suda \& Fujimoto \cite{suda10}, Karakas \cite{karakas10})

However, the observational evidence of this low-metallicity phenomenon is still scarce.  Observations of increasingly larger  samples of
Local Group PNe helped to analyze the occurrence of the TDU. In Figure~\ref{fig_TDU}
we show Ne/O, S/O and Ar/O versus O/H of several PN populations in dIrrs, plotted with the average oxygen abundance of the host galaxy HII 
regions (the vertical lines, see figure label).

The occurrence of the TDU  in PNe is identified by: {\em i)} higher O/H with respect to the corresponding ratio of HII regions; and 
{\em ii)} different Ne/O, S/O and Ar/O ratios with respect to the HII regions. Neon, sulphur, and
argon are dredged-up less efficiently than oxygen, in most cases.  We note PNe and HII regions have similar O/H ratios 
at high
metallicity, while at low metallicity, in particular when 
12+log(O/H)$<$ 7.7-7.8,  many PNe are richer in oxygen than the present-time ISM (in Fig.3,
these are found at the right of the average HII regions metallicity).
This does not occur for all the metal poor PNe, probably because the TDU is related not only to the
metal content, but also to the stellar mass and to other parameters such as rotation or magnetic fields.

A similar analysis can not be done in dwarf spheroidal galaxies, due to the absence of a HII region population. 
Useful diagrams that identify the presence
of the TDU in PNe belonging to dwarf spheroidals have been developed
by Richer \& McCall (\cite{rm08}). For example, the diagrams showing Ne/H vs O/H, or N/Ne vs Ne/O, can give useful hints of the TDU occurrence.
Richer \& McCall found a group of PNe having low Ne/O and high nitrogen
(which is usually a signature of massive progenitors),  indicating that,
occasionally, oxygen is produced during the evolution of the stellar
progenitors even in galaxies where star formation is not present.

\subsection{The chemical evolution of dwarf galaxies}

The large amount of information available for nearby dwarf galaxies,
such as their gas and star content, their metallicity distribution, and their star formation
history, allows us to use them to test chemical evolution models.
As an example, for IC10 Magrini \& Gon\c calves (\cite{mg09}) obtained
the composition of a sample of HII regions and PNe.
They found a very limited  enrichment in the metal content of IC10 in the time elapsed
between the birth of PN progenitors and the present-time. Yin et al (\cite{yin10})
compared these results  with a series of chemical evolutionary models in which they varied the star
formation history (from continuous to burst) and included different treatments and assumptions for galactic winds, namely respectively
a model with no winds at all, one with standard wind treatment, where the wind has the same chemical composition of the ISM, and finally a model with 
metal-enhanced winds.

Both models without wind and those with standard wind (i.e., stripping both H and
metals from the galaxy) disagree with the observational data, contrary to the model with a metal-enhanced wind, which  resulted to be in better
agreement with the observations. The best model to fit current data sets consists of a continuous wind that develops
from the first burst of star formation in the galaxy, and that strips the galaxy mostly its heavy
elements.







\begin{thebibliography}{}
\bibitem[2011]{bresolin11} Bresolin, F.\ 2011, \textit{ApJ}, 730, 129
\bibitem[1997]{chiappini97} Chiappini, C., Matteucci, F., \& Gratton, R.\ 1997, \textit{ApJ}, 477, 765
\bibitem[2001]{chiappini01} Chiappini, C., Matteucci, F., \& Romano, D.\ 2001, \textit{ApJ}, 554, 1044
\bibitem[2006]{crockett06} Crockett, N. R., Garnett, D. R., Massey, P., Jacoby, G.\ 2006, \textit{ApJ}, 637, 741
\bibitem[1991]{dopita91} Dopita, M.~A., \& Meatheringham, S.~J.\ 1991, \textit{ApJ}, 367, 115
\bibitem[2001]{freedman01} Freedman, W.~L., et al.\ 2001, \textit{ApJ}, 553, 47
\bibitem[2004]{galleti04} Galleti, S., Bellazzini, M., \& Ferraro, F.~R.\ 2004, \textit{A\&A}, 423, 925
\bibitem[1987]{gs87} Garnett, D.~R., \& Shields, G.~A.\ 1987, \textit{ApJ}, 317, 82
\bibitem[2007]{goncalves07} Gon{\c c}alves, D.~R., Magrini, L., Leisy, P., \& Corradi, R.~L.~M.\ 2007, \textit{MNRAS}, 375, 715
\bibitem[2008]{goncalves08} Gon{\c c}alves, D.~R., Magrini, L., Munari, U., Corradi, R.~L.~M., \& Costa, R.~D.~D.\ 2008, \textit{MNRAS}, 391, L84
\bibitem[2011]{goncalves11} Gon{\c c}alves, D.~R., Magrini, L., Martins, C. Quireza, A. Teodorescu, 2011,
\textit{MNRAS}, in press
\bibitem[2009]{hm09} Hern{\'a}ndez-Mart{\'{\i}}nez, L., Pe{\~n}a, M., Carigi, L., \& Garc{\'{\i}}a-Rojas, J.\ 2009, \textit{A\&A}, 505, 1027
\bibitem[2010]{karakas10} Karakas, A.~I.\ 2010, \textit{MNRAS}, 403, 1413
\bibitem[2005]{kniazev05} Kniazev, A.~Y., Grebel, E.~K., Pustilnik, S.~A., Pramskij, A.~G., \& Zucker, D.~B.\ 2005, \textit{AJ}, 130, 1558
\bibitem[2007]{kniazev07} Kniazev, A.~Y., Grebel, E.~K., Pustilnik, S.~A., \& Pramskij, A.~G.\ 2007, \textit{A\&A}, 468, 121
\bibitem[2008a]{kniazev08a} Kniazev, A.~Y., et al.\ 2008a, \textit{MNRAS}, 388, 1667
\bibitem[2008]{kniazev08} Kniazev, A.~Y., Pustilnik, S.~A., \& Zucker, D.~B.\ 2008b, \textit{MNRAS}, 384, 1045
\bibitem[2006]{izotov06} Izotov, Y.~I., Stasi{\'n}ska, G., Meynet, G., Guseva, N.~G., \& Thuan, T.~X.\ 2006, \textit{A\&A}, 448, 955
\bibitem[2011]{lepine11} Lepine, J.~R.~D., et al.\ 2011, arXiv:1106.3137
\bibitem[2004]{magrini04} Magrini, L., Perinotto, M., Mampaso, A., Corradi, R. L. M.\ 2004, \textit{A\&A}, 426, 779
\bibitem[2005]{magrini05} Magrini, L., Leisy, P., Corradi, R.~L.~M., Perinotto, M., Mampaso, A., \& V{\'{\i}}lchez, J.~M.\ 2005, \textit{A\&A}, 443, 115
\bibitem[2007a]{magrini07a} Magrini, L., Vilchez, J. M., Mampaso, A., Corradi, R. L. M., Leisy, P, 2007a, \textit{A\&A}, 470, 865
\bibitem[2007b]{magrini07b} Magrini, L., Corbelli, E., Galli, D. \ 2007b, \textit{A\&A}, 470, 843
\bibitem[2009]{mg09} Magrini, L., \& Gon{\c c}alves, D.~R.\ 2009, \textit{MNRAS}, 398, 280
\bibitem[2009]{magrini09} Magrini, L., Stanghellini, L., \& Villaver, E.\ 2009, ApJ, 696, 729
\bibitem[2010]{magrini10} Magrini, L., Stanghellini, L., Corbelli, E., Galli, D., \& Villaver, E.\ 2010, \textit{A\&A}, 512, A63
\bibitem[1997]{molla97} Moll\'a, M., Ferrini, F., \& Diaz, A.~I.\ 1997, ApJ, 475, 519
\bibitem[2005]{molla05} Moll{\'a}, M., \& D{\'{\i}}az, A.~I.\ 2005, \textit{MNRAS}, 358, 521
\bibitem[2007]{pena07} Pe{\~n}a, M., Stasi{\'n}ska, G., \& Richer, M.~G.\ 2007, \textit{A\&A}, 476, 745
\bibitem[2011]{rahimi11} Rahimi, A., Kawata, D., Prieto, C.~A., Brook, C.~B., Gibson, B.~K., \& Kiessling, A.\ 2011, \textit{MNRAS}, 415, 1469
\bibitem[2008]{rm08} Richer, M.~G., \& McCall, M.~L.\ 2008, \textit{ApJ}, 684, 1190
\bibitem[2008]{rs08} Rosolowsky, E., Simon, J. D., 2008, \textit{ApJ}, 675, 1213
\bibitem[2008]{rubin08} Rubin, R.~H., et al.\ 2008, \textit{MNRAS}, 387, 45
\bibitem[2006]{sara06} Sarajedini, A., Barker, M.~K., Geisler, D., Harding, P., \& Schommer, R.\ 2006, \textit{AJ}, 132, 1361
\bibitem[2009]{stanghellini08} Stanghellini, L.\ 2009, IAU Symposium, 256, 421
\bibitem[2010]{st10} Stanghellini, L., Magrini, L., Villaver, E., \& Galli, D.\ 2010, \textit{A\&A}, 521, A3
\bibitem[2010]{suda10} Suda, T., \& Fujimoto, M.~Y.\ 2010, \textit{MNRAS}, 405, 177
\bibitem[2010]{yin10} Yin, J., Magrini, L., Matteucci, F., Lanfranchi, G.~A., Gon{\c c}alves, D.~R., \& Costa, R.~D.~D.\ 2010, \textit{A\&A}, 520, A55





\end{thebibliography}
\end{document}